\documentclass[aps,12pt,a4paper,showpacs,byrevtex]{revtex4}
\usepackage{amsmath}
\usepackage{bm}
\usepackage{amsfonts}
\begin{document}

\baselineskip 15pt

\title{Wave packets of a harmonic oscillator with
various\\ degrees of rigidity}\thanks{published in Chin. J. Phys.
(Taipei) {\textbf 40} (2002) 387-394.}



\author{Qiong-Gui Lin}
\email[]{qg_lin@163.net} \email[]{qg_lin@sina.com}

%\thanks{}

\affiliation{Department of Physics, Sun Yat-Sen University, Guangzhou
510275, People's  Republic of China} \affiliation{China Center of
Advanced Science and Technology (World Laboratory), P. O. Box 8730,
Beijing 100080, People's Republic of China}
\thanks{not for correspondence}

%\date{\today}

\begin{abstract}
{\normalsize The time evolution of wave packets in a harmonic
oscillator potential is studied. Some new results for the most
general case are obtained. A natural number, called ``degree of
rigidity'', is introduced to describe qualitatively how much the
shape of a wave packet is changed with time. Two classes of wave
packets with an arbitrarily given degree of rigidity are presented.}
\end{abstract}

\maketitle


%\newpage

\section{\label{s1} Introduction}

In recent years, the time evolution of localized wave packets of
various quantum-mechanical systems has been widely discussed in the
literature (see Ref. \cite{ajp96} for a review). As for wave packets
of the harmonic oscillator, the subject is as old as quantum
mechanics itself. Though some aspects of the Gaussian wave packets
(displaced or squeezed ground state) are still being discussed in the
current literature \cite{ejp99.1,epj99.2}, the problem was
essentially solved in the very early years
\cite{sch1926,ken1927,schiff}. In the 1950's, several authors studied
more general displaced number states and found that they keep their
shape unchanged while their center oscillates like a classical
particle \cite{pr54,pr56,polon54,ajp59}. Displaced and squeezed
number states of special forms were also considered in some of these
papers \cite{pr56,polon54}; it turns out that their width is also
oscillating, so the shapes of such wave packets change with time
apparently. Displaced and squeezed number states of the most general
form were studied in a recent paper \cite{nieto97}. This may
represent the most general case where the time dependence of the wave
packets could be worked out explicitly.

The questions we are concerned with are: First, if the time
dependence of the wave packet cannot be obtained explicitly, what
conclusions could be made in regard to its time evolution? Second, if
the shape of the wave packet changes with time, how can we describe
the level of changes for different cases? For the first question, it
is known that the center of the wave packet moves like a classical
particle while its width pulsates. However, one can find wave packets
whose width keeps unchanged but whose shape still changes with time.
To describe the change of the shape in the general case we introduce
general-order moments of $x$ about its center, defined as the mean
value of $(x-\bar x_t)^K$ and denoted by $Q_K(t)$, where $\bar x_t$
is the mean value of $x$ (which represents the center of the wave
packet) and $K$ a natural number. Differential equations for these
moments are established for general $K$, and are solved for the most
general initial conditions up to $K=4$. They are essentially
oscillating, but higher-order ones involve higher frequencies and
thus their time dependence is more complicated. Wave packets could be
found whose $Q_K(t)$ are time independent up to $K=4$. Obviously the
shape of such wave packets changes with time less than those that
only keep their width [represented by $\sqrt{Q_2(t)}$] unchanged. At
this stage the answer to the second question is becoming clear. We
introduce a natural number called the ``degree of rigidity'' for a
wave packet: If $Q_K(t)$ is time independent for $K=2,3,\ldots, 2N$
[note that $Q_0(t)\equiv 1$ and $Q_1(t)\equiv 0$] but not for
$K=2N+2$ (for $K=2N+1$ it may or may not be time independent), we say
that the wave packet has degree of rigidity $N$. It is obvious that
the larger $N$ is, the less the shape changes with time. The shape of
a wave packet with a time-dependent width changes with time
apparently and thus has no rigidity. One with a constant width but
with all higher-order moments time dependent has the ``ground''
degree of rigidity 1. On the other hand, if $Q_K(t)$ is time
independent for all $K$, then the degree of rigidity for such a wave
packet is infinity. In other words, its shape is perfectly rigid.
Typical examples of such wave packets are displaced number states.
Two classes of wave packets with an arbitrarily given degree of
rigidity are presented.

\section{\label{s2} Some general results}

Consider localized wave packets of the harmonic oscillator in one
dimension, whose time evolution is governed by the Hamiltonian
\begin{equation}\label{1}
H={p^2\over 2\mu}+{1\over 2}\mu\omega^2 x^2.
\end{equation}
The normalized initial wave function $\psi(x,0)$ at $t=0$ will be
denoted by $\psi_0$ for convenience. Localization means that the mean
value of $x^k p^l$ in $\psi_0$ is finite for any nonnegative $k$ and
$l$. The wave function $\psi(x,t)$ at time $t$ will be denoted by
$\psi_t$. The mean values of $x$ and $p$ at time $t$ are
\begin{equation}\label{2}
\bar x_t=(\psi_t, x\psi_t),\quad \bar p_t=(\psi_t, p\psi_t).
\end{equation}
Using the Schr\"odinger equation it is easy to show that
\begin{equation}\label{3}
\dot{\bar{x}}_t=\bar p_t/\mu, \quad \dot{\bar{p}}_t=-\mu\omega^2 \bar
x_t.
\end{equation}
The solution of these equations is
\begin{equation}\label{4}
\bar x_t=\bar x_0\cos\omega t+{\bar p_0\over\mu\omega}\sin\omega t,
\quad \bar p_t=\bar p_0\cos\omega t-{\mu\omega\bar x_0}\sin\omega t.
\end{equation}
The solution means that the center of an arbitrary wave packet moves
like a classical particle. This is a well known result.

To study the change with time of the shape of the wave packet we
consider the mean value of $(x-\bar x_t)^K$ in $\psi_t$ where $K$ is
a natural number. These mean values are denoted by $Q_K(t)$ in the
following. If the $Q_K(t)$'s are time independent for all $K$, then
the wave packet obviously keeps its shape unchanged while it is
moving, and this is the case for displaced number states
\cite{pr54,pr56,polon54,ajp59}. In order to examine their time
dependence for a general wave packet, we define the real quantities
\begin{equation}\label{5}
R_{kl}(t)={\frac 12}\bm( \psi_t, \{ (x-\bar x_t)^k, (p-\bar p_t)^l
\}\psi_t \bm), \quad S_{kl}(t)={\frac 1{2{\mathrm i}}}\bm( \psi_t,
[(x-\bar x_t)^k, (p-\bar p_t)^l ]\psi_t \bm),
\end{equation}
where $k$ and $l$ are nonnegative integers, and $\{F, G\}=FG+GF$,
$[F, G]=FG-GF$ for any operator $F$ and $G$. Obviously
$R_{K0}(t)=Q_K(t)$. In the following we also denote $R_{0K}(t)$ by
$P_K(t)$. It is straightforward to show that they satisfy the
following equations
\begin{subequations}\label{6}
\begin{equation}\label{6a}
\dot R_{kl}(t)={k\over\mu}R_{k-1,l+1}(t)-l\mu\omega^2 R_{k+1,l-1}(t)
+{\hbar\over 2\mu}k(k-1) S_{k-2,l}(t)-{\hbar\mu\omega^2\over2} l(l-1)
S_{k,l-2}(t),
\end{equation}
\begin{equation}\label{6b}
\dot S_{kl}(t)={k\over\mu}S_{k-1,l+1}(t)-l\mu\omega^2 S_{k+1,l-1}(t)
-{\hbar\over 2\mu}k(k-1) R_{k-2,l}(t)+{\hbar\mu\omega^2\over2} l(l-1)
R_{k,l-2}(t).
\end{equation}
\end{subequations}
Note that by definition we have $S_{k0}(t)=S_{0k}(t)=0$ for all $k$.
We are mainly interested in $R_{kl}(t)$, especially $Q_K(t)$. It is
easy to realize that the equations for the subset
$\{R_{kl}(t)|k+l=K\}$ close among themselves provided that the
lower-order subset $\{S_{kl}(t) |k+l=K-2\}$ has been obtained. The
equations for the subset $\{S_{kl}(t) |k+l=K\}$ also close among
themselves, provided that the lower-order subset
$\{R_{kl}(t)|k+l=K-2\}$ is known. Now for $k=l=0$ we have
$R_{00}(t)=1$ and $S_{00}(t)=0$ by definition. Similarly for $k+l=1$
we have $R_{10}(t)=R_{01}(t)=S_{10}(t)=S_{01}(t)=0$. These enable us
to solve the cases with $k+l\ge 2$.

The case $k+l=2$ is of essential interest, because it involves the
width of the wave packet. By definition we have
$S_{20}(t)=S_{02}(t)=0$, and $S_{11}(t)=\hbar/2$. These will be
useful when the case $k+l=4$ is considered. For the moment we are
interested in the remaining three. It is not difficult to find that
\begin{subequations}\label{7}
\begin{eqnarray}
&&Q_2(t)={\mu^2\omega^2 Q_2(0)+P_2(0)\over 2\mu^2\omega^2}+
{\mu^2\omega^2 Q_2(0)-P_2(0)\over 2\mu^2\omega^2}\cos 2\omega t
+{R_{11}(0)\over\mu\omega}\sin 2\omega t, \label{7a}\\
&&P_2(t)={\mu^2\omega^2 Q_2(0)+P_2(0)\over 2}-{\mu^2\omega^2 Q_2(0)-
P_2(0) \over 2}\cos 2\omega t -{\mu\omega R_{11}(0)}\sin 2\omega t,
\label{7b}\\
&&R_{11}(t)=R_{11}(0)\cos 2\omega t- {\mu^2\omega^2
Q_2(0)-P_2(0)\over 2\mu\omega}\sin 2\omega t.\label{7c}
\end{eqnarray}
\end{subequations}
The width of the wave packet at time $t$ is characterized by the
quantity $\Delta_t x=\sqrt{Q_2(t)}$, and that in the momentum space
by $\Delta_t p=\sqrt{P_2(t)}$. Thus the width of the wave packet
oscillates with frequency $2\omega$, just like that for a squeezed
ground state. It can be shown that both $Q_2(t)$ and $P_2(t)$ are
positive as they should be. The uncertainty product $\Delta_t x
\Delta_t p$ is also oscillating. These conclusions (for a general
wave packet) have been discussed in some different way previously
\cite{pr56,polon54}, so we will not go into more details. A relation
that seems not to be emphasized in the literature is
\begin{equation}\label{8}
\mu^2\omega^2(\Delta_t x)^2+(\Delta_t p)^2=\mu^2\omega^2 (\Delta_0
x)^2+(\Delta_0 p)^2.
\end{equation}
According to this result, $\Delta_t x$ reaches its minimum when
$\Delta_t p$ reaches its maximum and vice versa.

When $k+l=3$, all the $S_{kl}(t)$ can be found to be zero. The
equations for the $R_{kl}(t)$ can be solved without much difficulty.
They are all linear combinations of $\{ \sin\omega t, \cos\omega t,
\sin3\omega t, \cos3\omega t\}$. Since the results are lengthy and
not important for further discussions we will not write them down. We
just point out that they all vanish if the corresponding initial
values $R_{kl}(0)$ are all zero.

When $k+l=4$, it can be found that $S_{40}(t)=S_{04}(t)=0$,
$S_{31}(t)=3\hbar Q_2(t)/2$, $S_{13}(t)=3\hbar P_2(t)/2$, and
$S_{22}(t)=2\hbar R_{11}(t)$. However, the results for the
$R_{kl}(t)$ are rather lengthy, they are linear combinations of $\{1,
\sin2\omega t, \cos2\omega t, \sin4\omega t, \cos4\omega t\}$. We
only write down one of them here:
\begin{eqnarray}\label{9}
Q_4(t)&=&{3\mu^4\omega^4 Q_4(0)+3 P_4(0)+6\mu^2\omega^2 R_{22}(0)
+3\hbar^2\mu^2\omega^2\over 8\mu^4\omega^4} \nonumber\\
&+&{\mu^4\omega^4 Q_4(0)- P_4(0)\over 2\mu^4\omega^4}\cos2\omega t +
{R_{13}(0)+\mu^2\omega^2 R_{31}(0)\over \mu^3\omega^3}\sin 2\omega t
\nonumber \\
&+&{\mu^4\omega^4 Q_4(0)+ P_4(0)-6\mu^2\omega^2 R_{22}(0)
-3\hbar^2\mu^2\omega^2\over 8\mu^4\omega^4}\cos4\omega t \nonumber \\
&-&{R_{13}(0) -\mu^2\omega^2 R_{31}(0)\over 2\mu^3\omega^3}\sin
4\omega t.
\end{eqnarray}
We see that higher-order moments about the center have more
complicated time dependence, but they are essentially oscillating and
hence are periodic functions. This is also true for still higher
ones. Indeed, $\bar x_t$, $\bar p_t$ and $\psi_t$ are all periodic,
and so are $R_{kl}(t)$ and $S_{kl}(t)$.

For larger values of $k+l$, the solutions are more difficult to find.
We will not proceed further in this respect.

\section{\label{s3} Simplified results for special cases}

The results obtained in Sec. \ref{s2} are rather complicated. In the
following sections we will discuss the results for special cases with
the initial condition
\begin{subequations}\label{10}
\begin{equation}\label{10a}
\psi_0=\psi(x,0)=\varphi(x-x_0)\mathrm e^{\mathrm i p_0x/\hbar},
\end{equation}
where $x_0$ and $p_0$ are real constants, and $\varphi(x)$ has
definite parity, namely
\begin{equation}\label{10b}
\varphi(-x)=\pm\varphi(x).
\end{equation}
\end{subequations}
In this initial state it is not difficult to show that
\begin{equation}\label{11}
\bar x_0=x_0,\quad \bar p_0=p_0,
\end{equation}
and
\begin{equation}\label{12}
R_{kl}(0)=0,\quad S_{kl}(0)=0,\quad k+l=1,3,5,\ldots.
\end{equation}
The latter could be derived from the useful relations
\begin{eqnarray}\label{13}
&&\bm(\psi_0, (x-x_0)^k (p-p_0)^l\psi_0\bm)=\bm(\varphi(x), x^k
p^l \varphi(x)\bm),\nonumber \\
&&\bm(\psi_0, (p-p_0)^l (x-x_0)^k\psi_0\bm)=\bm(\varphi(x), p^l x^k
\varphi(x)\bm),\nonumber \\
&&k,l=0,1,2,\ldots,
\end{eqnarray}
which are easy to show.

The shape of the initial wave packet, $|\psi_0|^2$, is obviously
symmetric about the center $x_0$. This symmetry will be kept at later
times. To prove this it is sufficient to show that $Q_{2K-1}(t)=0$
for all natural numbers $K$. Indeed, we will show that
\begin{equation}\label{14}
R_{kl}(t)=0,\quad S_{kl}(t)=0,\quad k+l=1,3,5,\ldots.
\end{equation}
For $k+l=1$ it is true as given in Sec. \ref{s2}. Now suppose that it
is true for $k+l=2K-1$, and consider the case with $k+l=2K+1$. It is
easy to realize that the equations for the subset
$\{R_{kl}(t)|k+l=2K+1\}$ close among themselves and are all
homogeneous because all of the $S_{kl}(t)$ in the subset
$\{S_{kl}(t)|k+l=2K-1\}$ vanish (according to the assumption). Since
the initial conditions are all homogeneous too, the solutions are
obviously $R_{kl}(t)=0$ ($k+l=2K+1$). Similarly, $S_{kl}(t)=0$ for
$k+l=2K+1$ as well. The case with $k+l=3$ has been explicitly
calculated, and the result is consistent with this general
conclusion, as pointed out in Sec. \ref{s2}.

If $\varphi(x)$ is real, we have another useful consequence:
\begin{equation}\label{15}
R_{kl}(0)=0,\quad k,l=1,3,5,\ldots.
\end{equation}
As a result the solutions $R_{kl}(t)$ with $k+l=2K$ are also
simplified. For example, the sine terms in Eqs. (\ref{7a}),
(\ref{7b}), and (\ref{9}) all vanish. However, that $\varphi(x)$ is
real is just a sufficient condition, not a necessary one for Eq.
(\ref{15}).

Now if $\varphi(x)$ in the initial state satisfies the condition
\begin{equation}\label{16}
\bm(\varphi(x),(xp+px)\varphi(x)\bm)=0,\quad
\bm(\varphi(x),\mu^2\omega^2 x^2\varphi(x)\bm)=
\bm(\varphi(x),p^2\varphi(x)\bm),
\end{equation}
we have [cf. Eq. (\ref{13})]
$$
R_{11}(0)=0,\quad \mu^2\omega^2 Q_2(0)=P_2(0),
$$
and from Eq. (\ref{7}) we obtain
$$
Q_2(t)=Q_2(0),\quad P_2(t)=P_2(0),\quad R_{11}(t)=0.
$$
That is, the width of the wave packet keeps unchanged. The second
condition in Eq. (\ref{16}) means that the kinetic energy and the
potential energy have the same mean value in $\varphi(x)$. It is easy
to realize that all the number states $\varphi_n(x)$ satisfy the
conditions (\ref{10b}) and (\ref{16}). However, there exist many
other functions that satisfy these conditions. For example,
\begin{subequations}\label{17}
\begin{equation}\label{17a}
\varphi^{\text{even}}(x)=\sum_{i=1}^\infty a_i\varphi_{2n_i}(x),
\quad 0\le n_1<n_2<\ldots, \quad \sum_{i=1}^\infty |a_i|^2=1
\end{equation}
and
\begin{equation}\label{17b}
\varphi^{\text{odd}}(x)=\sum_{i=1}^\infty b_i\varphi_{2n_i+1}(x),
\quad 0\le n_1<n_2<\ldots, \quad \sum_{i=1}^\infty |b_i|^2=1
\end{equation}
\end{subequations}
all have definite parity. And as long as $n_{i+1}-n_i\ge 2$ for all
$i$, they also satisfy the conditions in Eq. (\ref{16}).

However, the shapes of such wave packets change with time, though
their width keeps unchanged. The reason is that $Q_4(t)$, and in
general the higher-order moments, are still oscillating. The
conditions for $Q_4(t)$, $P_4(t)$ etc. to be time independent are
\begin{equation}\label{18}
R_{13}(0)=R_{31}(0)=0,\quad \mu^4\omega^4 Q_4(0)=P_4(0),\quad
2\mu^2\omega^2 Q_4(0)-6R_{22}(0)= 3\hbar^2.
\end{equation}
If $\varphi(x)=\varphi_n(x)$, a number state, then it can be shown
that these conditions are satisfied, as expected. However, there
exist many other functions that satisfy these conditions. In fact,
the two classes of functions given in Eq. (\ref{17}) do if
$n_{i+1}-n_i\ge 3$ for all $i$. In this case Eq. (\ref{16}) is of
course satisfied too. Since now $Q_4(t)$ is also time independent,
the shapes of the wave packets changes with time less than the case
where only the width is kept unchanged. In other words, their shapes
are more rigid.

In general, the higher-order moments, say $Q_6(t)$, of the above wave
packets will still be time dependent. It will become more and more
difficult to discuss the problem in the above manner. Since we are
not going to find the most general results for these higher-order
moments, we will proceed in a different way in the next section.

\section{\label{s4} Wave packets with various degrees of rigidity}

In the last section we have found wave packets whose width keeps
unchanged with time, and ones whose $Q_K(t)$ up to $K=4$ all keep
unchanged with time. Obviously the shape of the latter changes with
time less than the former. Now we introduce a natural number called
the ``degree of rigidity'' for a wave packet: If $Q_K(t)$ is time
independent for $K=2,3,\ldots, 2N$ [note that $Q_0(t)\equiv 1$ and
$Q_1(t)\equiv 0$] but not for $K=2N+2$ (the situation for $K=2N+1$ is
not important in this definition), we say that the wave packet has
degree of rigidity $N$. Thus the two cases mentioned above have
degrees of rigidity 1 and 2, respectively. It is obvious that the
larger $N$ is, the less the shape changes with time. If the width of
a wave packet changes with time, then its shape changes with time
apparently and thus has no rigidity. On the other hand, if $Q_K(t)$
is time independent for all $K$, then the degree of rigidity for such
a wave packet is infinity. In other words, its shape is perfectly
rigid. Typical examples of such wave packets are displaced number
states.

In order to find wave packets with a  given degree of rigidity, we
consider the quantity
\begin{equation}\label{19}
W_{kl}(t)=\bm( \psi_t, (x-\bar x_t)^k (p-\bar p_t)^l \psi_t \bm)
=R_{kl}(t)+\mathrm i S_{kl}(t).
\end{equation}
Because $\psi_t=\mathrm e^{-\mathrm i Ht/\hbar}\psi_0$, we have
\begin{equation}\label{20}
W_{kl}(t)=\bm( \psi_0, (x_t-\bar x_t)^k (p_t-\bar p_t)^l \psi_0 \bm),
\end{equation}
where
\begin{subequations}\label{21}
\begin{equation}\label{21a}
x_t=\mathrm e^{\mathrm i Ht/\hbar}x\mathrm e^{-\mathrm i
Ht/\hbar}=x\cos\omega t+{p\over \mu\omega}\sin\omega t,
\end{equation}
\begin{equation}\label{21b}
p_t=\mathrm e^{\mathrm i Ht/\hbar}p\mathrm e^{-\mathrm i
Ht/\hbar}=p\cos\omega t-{\mu\omega x}\sin\omega t.
\end{equation}
\end{subequations}
Now we confine ourselves to initial states of the form (\ref{10}). On
account of Eqs. (\ref{4}) and (\ref{11}), we have
\begin{equation}\label{22}
W_{kl}(t)=\bm( \psi_0, [(x-x_0)\cos\omega
t+(\mu\omega)^{-1}(p-p_0)\sin\omega t]^k [(p-p_0)\cos\omega
t-{\mu\omega}(x-x_0)\sin\omega t]^l \psi_0 \bm).
\end{equation}
Note that for any function $f(x)$ we have $(p-p_0)[\mathrm e^{\mathrm
i p_0x/\hbar} f(x)]=\mathrm e^{\mathrm i p_0x/\hbar}pf(x)$, and that
the inner product is an integral over $x$, the above equation can be
simplified as
\begin{equation}\label{23}
W_{kl}(t)=\bm( \varphi(x), [x\cos\omega t+(\mu\omega)^{-1}p\sin\omega
t]^k (p\cos\omega t-{\mu\omega}x\sin\omega t)^l \varphi(x) \bm).
\end{equation}
Using Eqs. (\ref{21}) again this becomes
\begin{equation}\label{24}
W_{kl}(t)=\bm( \varphi_t(x), x^k p^l \varphi_t(x) \bm),
\end{equation}
where
\begin{equation}\label{25}
\varphi_t(x)=\mathrm e^{-\mathrm i Ht/\hbar}\varphi(x).
\end{equation}

If $\varphi(x)=\varphi_n(x)$, a number state, then
$\varphi_t(x)=\mathrm e^{-\mathrm i E_n t/\hbar}\varphi_n(x)$, and
\begin{equation}\label{26}
W_{kl}(t)=\bm( \varphi_n(x), x^k p^l \varphi_n(x)
\bm)=W_{kl}(0),\quad k,l=0,1,2,\ldots.
\end{equation}
Especially, $Q_K(t)=W_{K0}(t)=Q_K(0)$ for all $K$, so that the shape
of a displaced number state is perfectly rigid. This is a well-known
conclusion. Here we obtain the conclusion in a different and also
simple way.

Consider the function $\varphi^{\text{even}}(x)$ given in Eq.
(\ref{17a}), we have
\begin{equation}\label{27}
\varphi^{\text{even}}_t(x)=\sum_{i=1}^\infty a_i \exp(-\mathrm i
E_{2n_i} t/\hbar)\varphi_{2n_i}(x).
\end{equation}
Assuming that $n_{i+1}-n_i\ge N+1$. Because $x^kp^l\varphi_{2n_i}(x)$
is a linear combination of $\{ \varphi_{2n_i+k+l}(x),
\varphi_{2n_i+k+l-2}(x), \ldots, \varphi_{2n_i-(k+l)}(x) \}$, we have
for $k+l\le 2N$ that
\begin{equation}\label{28}
W_{kl}(t)=\bm( \varphi^{\text{even}}_t(x), x^k p^l
\varphi^{\text{even}}_t(x) \bm)=\sum_{i=1}^\infty |a_i|^2 \bm(
\varphi_{2n_i}(x), x^kp^l\varphi_{2n_i}(x) \bm)=W_{kl}(0).
\end{equation}
Especially
\begin{equation}\label{29}
Q_K(t)=Q_K(0),\quad K=2,3,\ldots,2N.
\end{equation}
The same result can be obtained for the function
$\varphi^{\text{odd}}(x)$ given in Eq. (\ref{17b}). [Remember that
$Q_K(t)=0$ for all odd $K$ as proved in Sec. \ref{s3}.] Thus, if
$n_{i+1}-n_i\ge N+1$, the two classes of functions given in Eq.
(\ref{17}) lead to wave packets [whose initial wave functions are
given by Eq. (\ref{10a})] that have degree of rigidity not less than
$N$. If some of the differences $n_{i+1}-n_i$ equals $N+1$, they will
in general have degree of rigidity $N$. Otherwise the degree of
rigidity will be greater than $N$. For large $N$, the shapes of these
wave packets are almost unchanged with time.

\section{\label{s5} Summary}

In this paper we studied the time evolution of a general wave packet
in the harmonic oscillator potential by examining the time dependence
of the various moments $Q_K(t)$ of $x$ about the center of the wave
packet. The differential equations for these objects are derived for
general $K$, and are solved for the most general initial conditions
up to $K=4$. Wave packets with constant width, and ones whose
$Q_K(t)$ up to $K=4$ are all constants, are discussed. These include
displaced number states as special cases. In general the shapes of
these wave packets still change with time because the higher order
moments are oscillating. A natural number $N$, called the degree of
rigidity, is introduced to describe qualitatively how much the shape
is changed with time. The larger $N$ is, the less the shape is
changed with time. Displaced number states are perfectly rigid and
have $N=\infty$. Two classes of wave packets with an arbitrarily
given degree of rigidity are given explicitly.

\begin{acknowledgments}
This work was supported by the National Natural Science Foundation of
the People's Republic of China, and by the Foundation of the Advanced
Research Center of Sun Yat-Sen University.
\end{acknowledgments}

%\newpage
\bibliography{wave}
\end{document}